\documentclass[12pt,fleqn]{article}
\usepackage{times}
\usepackage[T1]{fontenc}
\normalfont
\usepackage{fourier} 
\usepackage{graphicx}
\usepackage{wrapfig}
\usepackage{amssymb}
\usepackage{amsmath}
\usepackage{color}
\usepackage{cite}
\begin{document}
\baselineskip 0.6cm
\renewcommand{\thefootnote}{\#\arabic{footnote}} 
\setcounter{footnote}{0}
%
\begin{titlepage}
\begin{center}

\begin{flushright}
\end{flushright}


{\Large \bf 
Lepton flavor model with modular $A_4$ symmetry\\[1ex]
in large volume limit
}

\vskip 1.2cm

{
Takehiko Asaka$^1$,
Yongtae Heo$^{2}$,
and 
Takahiro Yoshida$^{2}$
}

\vskip 0.4cm

$^1${\em \small
  Department of Physics, Niigata University, Niigata 950-2181, Japan
}

$^2${\em \small
Graduate School of Science and Technology, Niigata University, 950-2181 Niigata, Japan}

\vskip 0.2cm

(September 25, 2020)

\vskip 2cm

\begin{abstract}
We consider the modular symmetry
associated with the compactification of extra dimensions
as the flavor symmetry on lepton sector.
Especially, we propose a model based on the modular $A_4$ symmetry
with three right-handed neutrinos and a gauge singlet Higgs,
which works well in the so-called large volume limit of the extra dimensions, 
{\textit{i.e.}}, $\mbox{Im}\tau \to \infty$ for a modulus $\tau$.
The right-handed neutrinos are introduced to realize the seesaw mechanism 
for tiny neutrino masses observed in oscillation experiments.
The vacuum expectation value of the singlet Higgs
gives the Majorana masses for right-handed neutrinos 
and the $\mu$-term for Higgs fields.
The model can explain the observed masses and mixing angles of 
neutrinos successfully.  
We find that one of the mixing angles should be in the range
$\sin^2 \theta_{23} \ge 0.58$.
Importantly, the CP violating phases of neutrinos are predicted in
the two restricted regions.   
One is that the Dirac phase is $\delta_{\rm CP} \simeq - 0.5 \pi$
and the Majorana phases are $\alpha_{21} \simeq 0$ and $\alpha_{31} \simeq \pi$.
The other is $\delta_{\rm CP} \simeq + 0.5 \pi$, $\alpha_{21} \simeq 2 \pi$ and $\alpha_{31} \simeq \pi$.
The effective neutrino mass in the neutrinoless double beta decay is 
found to be $m_{\rm eff} = 0.037$--0.047~eV.
These predictions will be tested in the future neutrino experiments.
\end{abstract}
\end{center}
\end{titlepage}

\section{Introduction}
Neutrino oscillation experiments have shown a distinct pattern of
flavor mixing among three active neutrinos.   This pattern can be 
naturally explained by the flavor symmetry imposed on leptonic fields.
Especially, the discrete symmetry, such as $S_3$, $A_4$ and so on,
can provide the successful fits to the oscillation data.
See, for example, the reviews~\cite{Altarelli:2010gt,Ishimori:2010au,Hernandez:2012ra,King:2013eh,Petcov:2017ggy}.

It has been pointed out that the modular symmetry in the torus compactification
of extra dimensions (to be precise, the quotient group of the modular group) can play a role of such a discrete flavor symmetry~\cite{Feruglio:2017spp}.
The modular symmetry restricts the structure of Yukawa coupling constants
in the superpotential together with the modular form, and the observed
mixing angles and masses of active neutrinos can be described by a limited
number of parameters together with the value of the modulus $\tau$.  
For this reason the models based
on this symmetry have been intensively discussed in the literature~\cite{Kobayashi:2018vbk,Penedo:2018nmg,
Criado:2018thu,Kobayashi:2018scp,Novichkov:2018ovf,deAnda:2018ecu,
Okada:2018yrn,Kobayashi:2018wkl,Novichkov:2018nkm,Novichkov:2018yse,
Ding:2019xna,Nomura:2019jxj,Novichkov:2019sqv,deMedeirosVarzielas:2019cyj,
Liu:2019khw,Okada:2019xqk,Kobayashi:2019mna,Ding:2019zxk,King:2019vhv,
Nomura:2019lnr,Criado:2019tzk,Kobayashi:2019xvz,Asaka:2019vev,
Chen:2019ewa,Gui-JunDing:2019wap,Zhang:2019ngf,Wang:2019ovr,
Kobayashi:2019uyt,Nomura:2019xsb,Kobayashi:2019gtp,Lu:2019vgm,
Wang:2019xbo,Okada:2020dmb,Ding:2020yen}.

Recently, Ref.~\cite{Chen:2019ewa} has shown that
the higher dimensional terms in the K\"ahler potential, which are 
consistent with the modular symmetry, might disturb the successful 
prediction of the neutrino oscillation observables.
A possibility to avoid this difficulty is taking the so-called large volume limit, which is represented as $\mbox{Im}\tau \to \infty$.  
In this case the form of the K\"ahler potential
is considered to be restricted in specific forms in the framework of string theories~\cite{Kaplunovsky:1995jw, Antoniadis:1994hg} (see also Ref.~~\cite{Chen:2019ewa}).
On the other hand, the flavor models based on the modular symmetries require
$\mbox{Im} \tau = {\cal O}(1)$ in order to obtain the realistic values of neutrino mixing angles 
and masses.

Under these observation, we would like to construct a model with the modular symmetry
which meets with the neutrino oscillation data even in the large volume limit.
As a first example, our model is based on the one with the modular $A_4$ symmetry
and the seesaw mechanism~\cite{Minkowski:1977sc,Yanagida:1979ws,Gell-Mann:1979ws,Glashow:1980wc,Mohapatra:1979ia} by three right-handed neutrinos
proposed in Ref.~\cite{Kobayashi:2018scp}.  
We then introduce a gauge singlet field $\hat S$ which is a triplet of the $A_4$ symmetry.
The VEV of $\hat S$ is crucial to explain the observed properties of neutrinos.
In addition, we assign the modular weights of the fields in order to be consistent
with the K\"ahler transformation of the theory, and the $\mu$ term of the Higgs fields
and the Majorana masses of right-handed neutrinos are generated by the VEV of $\hat S$.

In the following we first explain our model proposed in this letter
and present the mass matrices of leptons.
It is then demonstrated that the model can explain the observed mixing angles and 
mass squared differences of neutrinos.
Further, we show the predictions of the model on the unknown neutrino properties,
the mass ordering, the CP violating phases and the effective neutrino mass in the neutrinoless double beta decay.
Finally, we summarize the obtained results.

\section{Model}
Let us start to explain the model we propose in this letter.
It is based on the model I (a) in Ref.~\cite{Kobayashi:2018scp} 
which possesses the modular $A_4$ symmetry.
The masses of active neutrinos are
induced by the seesaw mechanism by right-handed neutrinos $\hat N^c$,
which is a triplet of the $A_4$ symmetry.
The lepton doublets $\hat L$ are also combined in a triplet of $A_4$
and right-handed charged lepton fields $\hat E^c_{1}$, $\hat E^c_2$, and $\hat E^c_3$
are assigned three distinct singlets 
$\bf 1$, $\bf 1''$, and $\bf 1'$, respectively.
The Higgs doublets $\hat H_u$ and $\hat H_d$ are both taken to be $\bf 1$.
See the field content in Tab.~\ref{Tab:Model}.
\begin{table}[t]
$$
\renewcommand{\arraystretch}{1.3}
\begin{array}{|c|c|c|c|c|c|c||c|}
\hline
& {\hat{L}} & {\hat{E_1^{c}}, \, \hat{E_2^{c}}, \,\hat{E_3^{c}}} & \hat{N^{c}} & \hat{H_{u}} & \hat{H_{d}} & \hat{S} & Y^{A_4}\\ \hline \hline
S U(2)_{L} & {\bf 2} & {\bf 1} & {\bf 1} & {\bf 2} & {\bf 2} & {\bf 1} & {\bf 1}\\ 
{A_{4}} & {\bf 3} & {\bf 1}, \, {\bf 1^{\prime \prime}}, \, {\bf 1^{\prime}} & {\bf 3} & {\bf 1} & {\bf 1} & {\bf 3} & {\bf 3}\\ 
\mbox{M.W.} & {-1} & {-1} & {-1} & {-1} & {-1} & {-1} & + 2\\ \hline
{R} & - & - & - & + & + & + & + \\ \hline
\end{array}
$$
\caption{The field content of the model.
We show the representation of each field under $SU(2)_L$ gauge group, modular $A_4$ group,
modular weight (M.W.), and $R$-parity ($R$).}

\label{Tab:Model}
\end{table}

Notice that the assignment of the modular weight for the Higgs fields is different from 
the one in Ref.~\cite{Kobayashi:2018scp}.  This is because we consider 
the model in the framework of supergravity and
require the invariance of the K\"ahler transformation
of the model.  In this case the modular weight of the superpotential is ${}-1$.
(See, for example, the discussion in Ref.~\cite{Kobayashi:2019xvz}.)

Further, we additionally introduce the gauge singlet Higgs field $\hat S$,
which transforms as $\bf 3$ under the modular $A_4$ group,
\begin{align}
  \hat S =
  \left(
    \begin{array}{c}
    \hat S_1 \\
    \hat S_2 \\
    \hat S_3
\end{array}
  \right) \,.
\end{align}
As mentioned in Introduction,
we would like to consider the large volume limit, {\it i.e.},
the imaginary part of the modulus $\tau$ to be $\mbox{Im}\tau \to \infty$.
In this case the $A_4$-triplet modular form takes the form as~\cite{Feruglio:2017spp}
\begin{align}
    Y^{A_4} &=
    \left(
    \begin{array}{l}
     1+12q+36q^2+12q^3+\cdots	\\
-6q^{\frac{1}{3}}(1+7q+8q^2+\cdots)	\\
-18q^{\frac{2}{3}}(1+2q+5q^2+\cdots)   
    \end{array}
    \right)
    \rightarrow\left(
        \begin{array}{c}
            1 \\ 0 \\ 0
        \end{array}
    \right) ~~~~~\mbox{for}~~~\mbox{Im}\tau \to \infty\,,
    \label{eq:MF}
\end{align}
where $q = \exp( 2 \pi i \tau)$.
This modular form gives too simple mass matrices of neutrinos, both Dirac and Majorana types,
to explain the observed pattern of neutrino mixing.
The VEV of $\hat S$ then induces the Majorana masses of right-handed neutrinos
with a complex structure, which leads to the successful fit to the observational data.
In addition, our assignment of the modular weights forbids the $\mu$ term of the Higgs fields,
which is also generated by the VEV of $\hat S$.

The explicit form of the superpotential which is relevant for the following discussion 
is given by
\begin{align}
    W &= k \, ( \hat S Y^{A_4} \hat H_u \hat H_d )_{\bf 1} 
    \nonumber \\
    &+ f_1 \,(\hat L Y^{A_4})_{\bf 1} \, \hat E^c_1 \, H_d 
    + f_2 \, (\hat L Y^{A_4})_{\bf 1'} \, \hat E^c_2 \, H_d 
    + f_3 \, (\hat L Y^{A_4})_{\bf 1"} \, \hat E^c_3 \, H_d 
    \nonumber \\
    &+ g_1 \, \big( (\hat L Y^{A_4})_{\bf 3s} \, \hat N^c \, \hat H_u \big)_{\bf 1}
    + g_2 \,  \big( (\hat L Y^{A_4})_{\bf 3a} \, \hat N^c \, \hat H_u \big)_{\bf 1}
    \nonumber \\
    &+ h_1 \, \big( (\hat S Y^{A_4})_{\bf 1} \hat N^c \, \hat N^c \big)_{\bf 1}
    + h_2 \, \big( (\hat S Y^{A_4})_{\bf 1'} \hat N^c \, \hat N^c \big)_{\bf 1}
    + h_3 \, \big( (\hat S Y^{A_4})_{\bf 1"} \hat N^c \, \hat N^c \big)_{\bf 1}
    \nonumber \\
    &+ h_4 \, \big( (\hat S Y^{A_4})_{\bf 3s} \hat N^c \, \hat N^c \big)_{\bf 1}
    + h_5 \, \big( (\hat S Y^{A_4})_{\bf 3a} \hat N^c \, N^c \big)_{\bf 1} \,.
\end{align}
where $k$, $f_{1,2,3}$, $g_{1,2}$, and $h_{1,2,3,4,5}$ are constant couplings,
and $\bf 3s$ or $\bf 3a$ denotes the symmetric triplet or anti-symmetric triplet
of $A_4$, respectively.

The VEV of $\hat S$ induces the $\mu$ term of the Higgs fields as
\begin{align}
    \mu = k \, S_1 \,,
\end{align}
where $S_I = \langle \hat S_I \rangle$ ($I=1,2,3$), which shows that
$S_1 \neq 0$ is needed for $\mu \neq 0$.
Remember that we assume the modular form (\ref{eq:MF}) in the large volume limit.
Here we have omitted the self couplings of $\hat S$, {\it i.e.},
the terms with $\hat S^3$ in the superpotential.
We do not specify how the $\hat S$ field obtains the VEV,
but simply assume its pattern since it is beyond the scope of this analysis.

First, we find that the case with $S_1 \neq 0$, $S_2=S_3=0$ is inconsistent with 
the neutrino oscillation data, and hence we discard this possibility.
We then consider the case in which the one VEV of three components to be zero for simplicity.
\begin{align}
    \langle \hat S \rangle
    =
    \left(
    \begin{array}{c}
    S_1 \\ S_2 \\ 0 
    \end{array}
    \right) 
    \,.
\end{align}
Since the $\mu$ term requires $S_1 \neq 0$, there are two possibilities
$S_2 = 0$ or $S_3 = 0$.
In this letter we show the result with $S_3=0$.

The mass matrix of charged leptons is given by
\begin{align}
    M_E &= \langle \hat H_d \rangle \,
    P_{ijk} \,
    \left(
    \begin{array}{ccc}
        f_1 & 0 & 0  \\
        0 & f_2 & 0  \\
        0 & 0 & f_3
    \end{array}
    \right) \,
    P_{ijk}^T\,.
\end{align}
Note that there are six possibilities
the assignment of $(\hat E^c_1, \hat E^c_2, \hat E^c_3)$
to three generations of right-handed charged leptons ($e^c$, $\mu^c$, $\tau^c$).
This connection is represented by the matrix $P_{ijk}$.
In this letter we present the results
with the case $E^c_1 = e^c$, $E_2^c = \mu^c$, and $E_3^c = \tau^c$.
In this case $P_{ijk} = \mbox{diag}(1,1,1)$.%
\footnote{
In general, there are two possibilities for the zero entry in the VEV of $\hat S$
and six possibilities for the assignment of $(\hat E^c_1, \hat E^c_2, \hat E^c_3)$.
One may then consider that there are twelve sets of mass matrices for leptonic fields.
We find, however, that the final mass matrix for active neutrinos through the seesaw matrix
is classified into only three types.  The details will be discussed elsewhere~\cite{AHY2}.}
Here and hereafter we work in the basis where
$M_E = \mbox{diag}(m_e, m_\mu, m_\tau)$.

The Dirac mass matrix of neutrinos takes the form
\begin{align}
    M_D = \langle \hat H_u \rangle \,
    \left(
    \renewcommand{\arraystretch}{1.2}
    \begin{array}{c c c}
    2 g_1 & 0 & 0 \\
    0 & 0 & - g_1 +g_2 \\
    0 & - g_1 -g_2 & 0
    \end{array}
    \right)\, 
    P_{ijk}^T\,,
\end{align}
and the Majorana mass matrix of right-handed neutrinos is
\begin{align}
    M_M = 
    \left(
    \renewcommand{\arraystretch}{1.2}
    \begin{array}{c c c}
    (h_1 + 4 h_4 ) S_1 & 0 & (h_2 + h_4 + h_5) S_2 \\
    0 & (h_2 - 2 h_4 - 2 h_5) S_2 & (h_1 -2 h_4) S_1 \\
    (h_2 + h_4 + h_5) S_2 & (h_1 - 2 h_4) S_1 & 0 
    \end{array}
    \right) \,.
\end{align}
Note that the coupling $h_3$ is irrelevant for the case with $S_3 =0$ under consideration. \footnote{Here we have redefined the couplings as $h_4 / 9 \to h_4$ and $h_5 / 6 \to h_5$. }

Then, the seesaw mechanism~\cite{Minkowski:1977sc,Yanagida:1979ws,Gell-Mann:1979ws,Glashow:1980wc,Mohapatra:1979ia} generates the mass matrix of active neutrinos as
\begin{align}
    M_\nu &
    = - M_D^T \, M_M^{-1} \, M_D
    =
    \Lambda 
    \left(
        \renewcommand{\arraystretch}{1.2}
        \begin{array}{ccc}
        1 & b_2 \, b_3 & b_3 \\
        b_2 \, b_3 & b_1 \, b_2 & b_1 \\
        b_3 & b_1  & b_3^2
        \end{array}
    \right) \,,
\end{align}
where
\begin{align}
    \Lambda =
    -
    \frac{
    4 g_1^2 (h_1-2 h_4)^2 \langle {\hat H}_u \rangle^2 \, S_1^2 
    }{
    (h_1 - 2 h_4)^2 (h_1 + 4 h_4) \, S_1^3
    +
    (h_2+h_4+h_5)^2 ( h_2 - 2 h_4 - 2 h_5) \, S_2^3 
    } \,,
\end{align}
and 
\begin{align}
    b_1 &=
    \frac{
    (g_1^2 -g_2^2)(h_1 + 4 h_4 )
    }{
    4 g_1^2 (h_1 - 2 h_4)
    } \,,
    \\[1ex]
    b_2 &= 
    - \frac{(g_1 + g_2)(h_2 - 2 h_4 - 2 h_5) \, S_2}
    {(g_1-g_2) ( h_1 - 2 h_4 ) \, S_1} \,,
    \\[1ex]
    b_3 &= \frac{
    (g_1 - g_2)(h_2 + h_4 + h_5) \, S_2 
    }{
    2 g_1 (h_1 - 2 h_4 ) \, S_1 
    } \,.
\end{align}

So far, we have ignored the explicit form of the K\"ahler potential and just 
considered the K\"ahler metric of lepton fields is diagonal.
The $Z_3$ symmetry is left in the large volume limit
with the modular form in Eq.~(\ref{eq:MF}),
which ensures the diagonal form of the K\"ahler metric.
Further, the corrections due to the VEV of $\hat S$ can be suppressed
by taking the VEV much smaller than the fundamental cutoff scale.
It should be noted that the field redefinition in the kinetic term can be absorbed by
the redefinition in the parameters $b_{1,2,3}$.   Thus, the predictions on 
the neutrino properties in this model, which will be discussed below,
do not change by the redefinition.

\section{Properties of neutrinos}
As explained above, the mass matrix of active neutrinos is described by 
one mass parameter $\Lambda$ and three complex coupling parameters
$b_1$, $b_2$, and $b_3$.   
We perform the numerical analysis to find the parameter region
in which the predictions of the neutrino mixing angles
$\theta_{12}$, $\theta_{23}$, $\theta_{13}$ 
and the mass squared differences are consistent with the $3 \sigma$ range
of the global analysis given in Ref.~\cite{nufit}.
Note that the mixing matrix of neutrinos is expressed as
\begin{align}
  U =
  \left( 
    \begin{array}{c c c}
      c_{12} c_{13} &
      s_{12} c_{13} &
      s_{13} e^{- i \delta_{\rm CP}} 
      \\
      - c_{23} s_{12} - s_{23} c_{12} s_{13} e^{i \delta_{\rm CP}} &
      c_{23} c_{12} - s_{23} s_{12} s_{13} e^{i \delta_{\rm CP}} &
      s_{23} c_{13} 
      \\
      s_{23} s_{12} - c_{23} c_{12} s_{13} e^{i \delta_{\rm CP}} &
      - s_{23} c_{12} - c_{23} s_{12} s_{13} e^{i \delta_{\rm CP}} &
      c_{23} c_{13}
    \end{array}
  \right)  
  \times
  \mbox{diag} 
  \left( 1 \,,~ e^{i \alpha_{21}/2} \,,~ e^{i \alpha_{31}/2}
  \right)\,,
\end{align}
where $s_{ij} = \sin \theta_{ij}$ and $c_{ij} = \cos \theta_{ij}$. $\delta_{\rm CP}$ and $\alpha_{21,31}$ are the Dirac and Majorana CP violating phases, respectively.
In addition we impose the cosmological constraint
on the sum of neutrino masses~\cite{Aghanim:2018eyx}
\begin{align}
    \sum_{i=1}^3 m_i \le 0.160~\mbox{eV} \,.
    \label{eq:sum_mi}    
\end{align}

\begin{figure}[t]
  \centerline{
  \includegraphics[width=7cm]{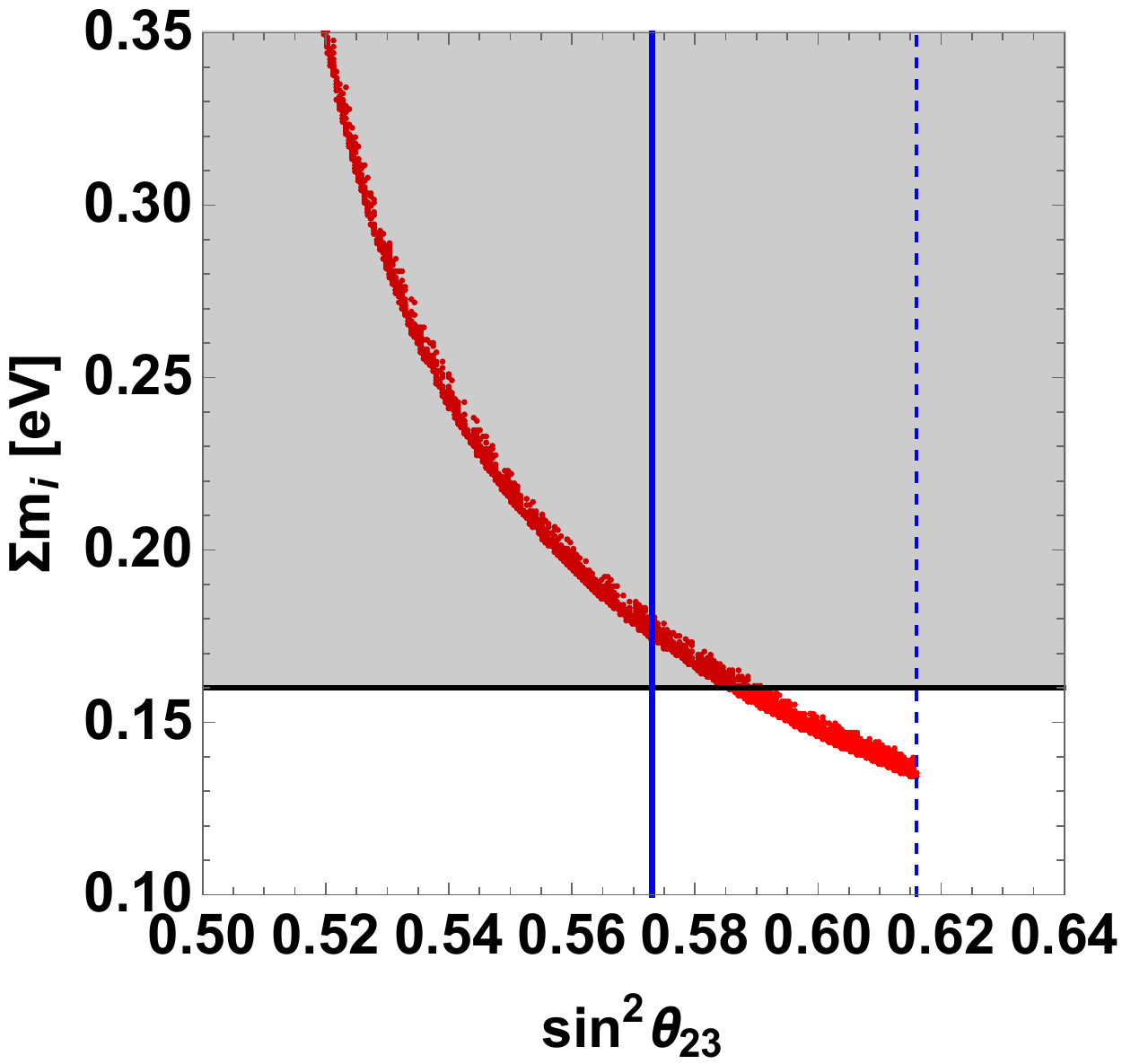}%
  }%
  \vspace{-2ex}
  \caption{
  	Predicted range of $\sum m_i$ and $\sin^2 \theta_{23}$ shown by red dots.
  	The solid and dashed vertical blue lines show the central and $3\sigma$
  	range of the observed value of $\sin^2 \theta_{23}$.
  	The colored region is excluded by the cosmological bound on $\sum m_i$ in Eq.~(\ref{eq:sum_mi}).
  }
  \label{Fig_S23_summi}
\end{figure}
First, we discuss the mass ordering of active neutrinos.
We find that the sum of masses in the inverted hierarchy (IH) case exceeds 
the bound in Eq.~(\ref{eq:sum_mi}), and only the normal hierarchy (NH) case
is allowed.  See Fig.~\ref{Fig_S23_summi}.
It is found that the model predicts $\sum m_i \ge 0.13$~eV and 
the bound in Eq.~(\ref{eq:sum_mi}) gives the lower bound on the 
$\sin^2 \theta_{23} \ge 0.58$.  Thus, a large value of
$\sin^2 \theta_{23}$ is required in this model.

\begin{figure}[t]
  \centerline{
  \includegraphics[width=7cm]{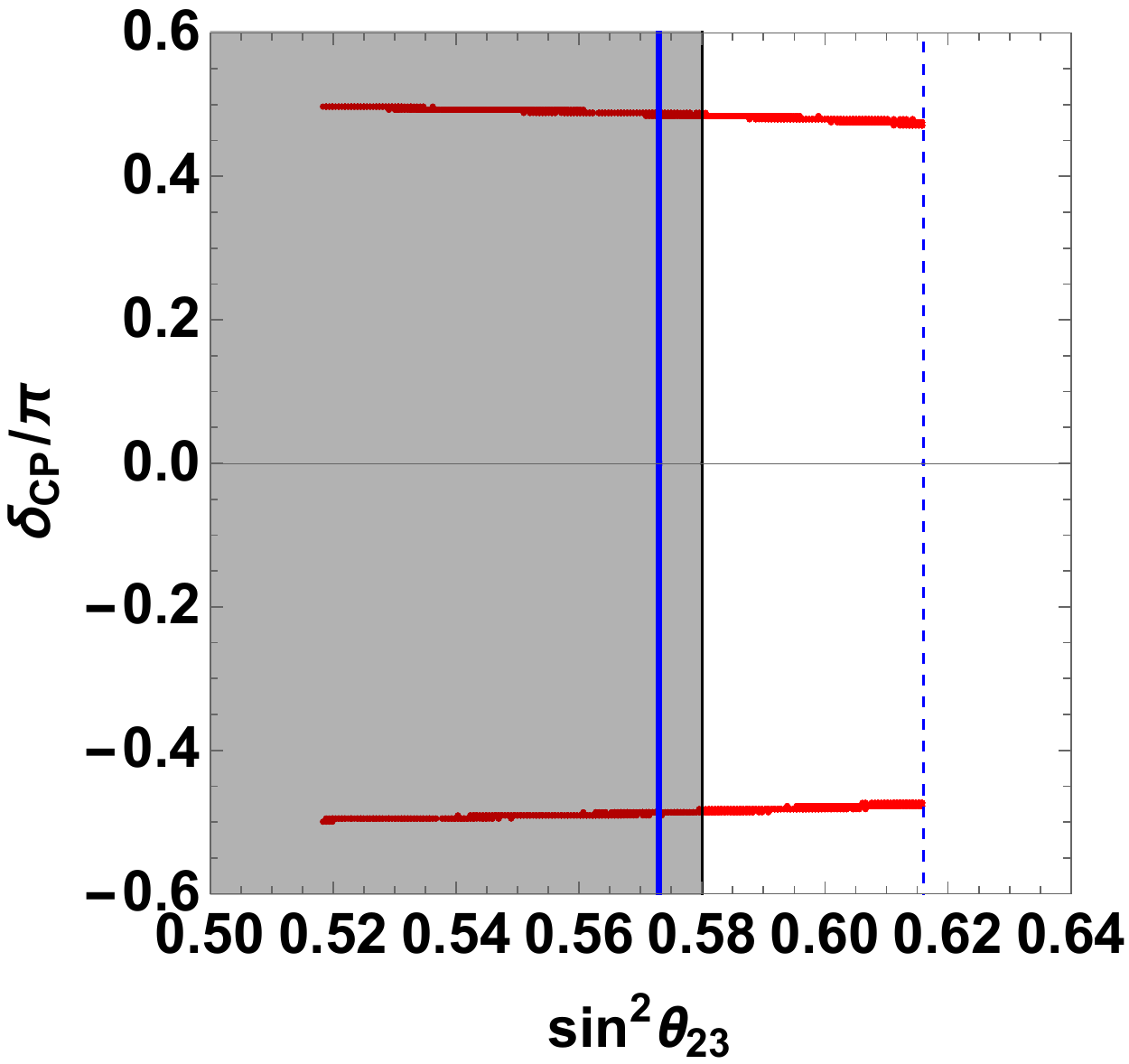}%
  }%
  \vspace{-2ex}
  \caption{
  	Predicted range of $\delta_{\rm CP}$ and $\sin^2 \theta_{23}$ shown by red dots.
  	The solid and dashed vertical blue lines show the central and $3\sigma$
  	range of the observed value of $\sin^2 \theta_{23}$.
  	The colored region is excluded by the cosmological bound on $\sum m_i$ in Eq.~(\ref{eq:sum_mi}).
  }
  \label{fig:Fig_S23_DEL}
  \vspace{0.5cm}
  \centerline{
  \includegraphics[width=7cm]{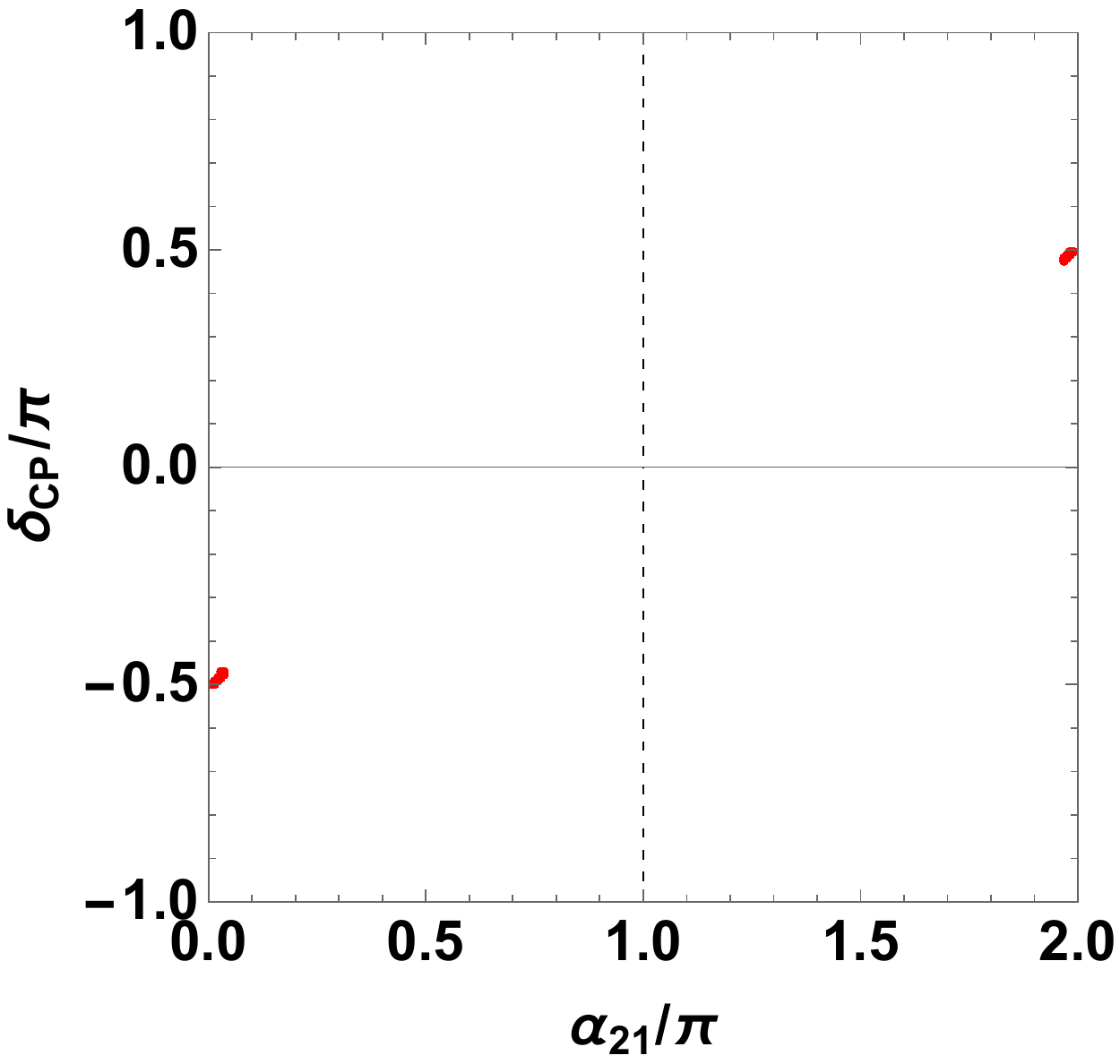}%
  \includegraphics[width=7cm]{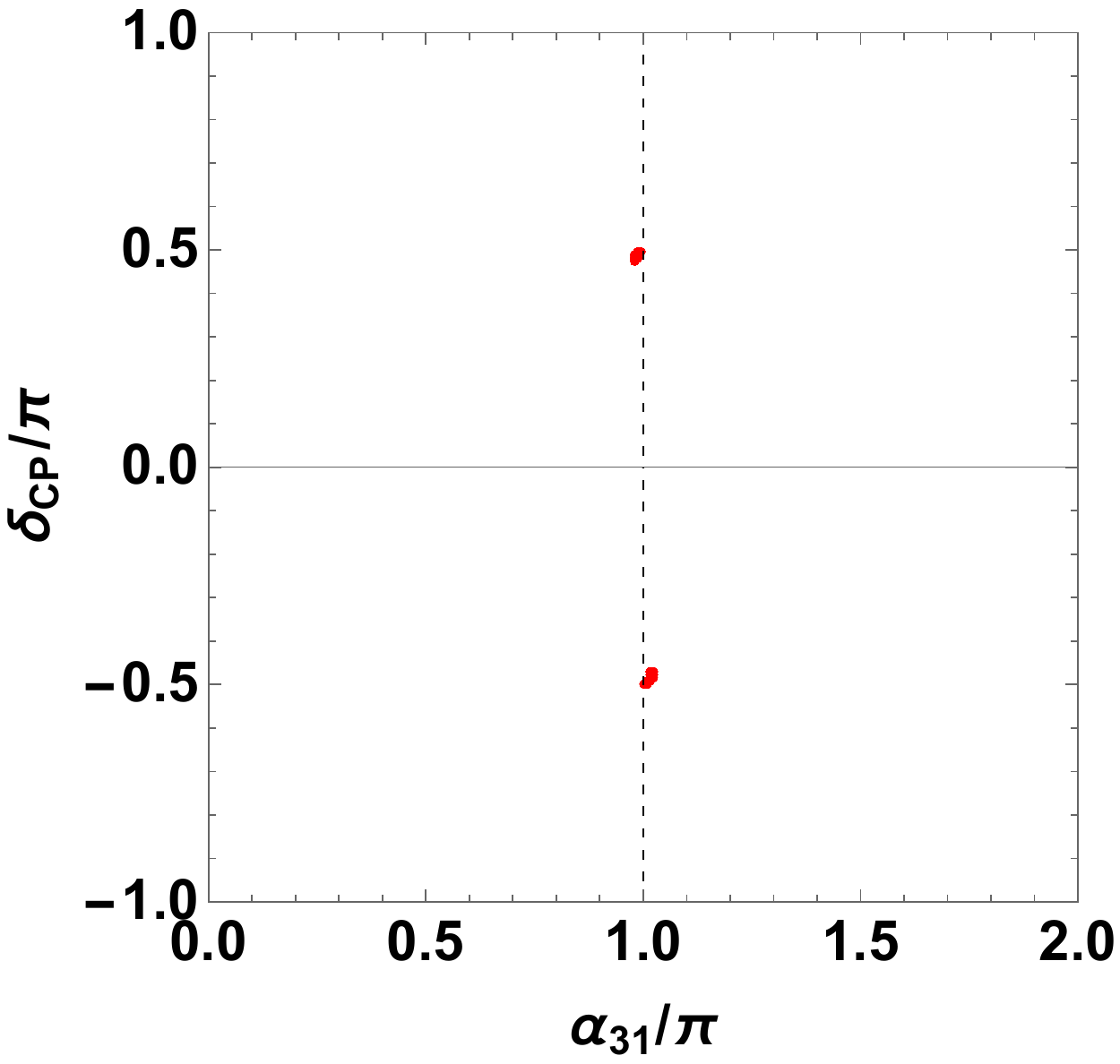}%
  }%
  \vspace{-2ex}
  \caption{
  	Predicted range of $\delta_{\rm CP}$ and $\alpha_{21}$ (left)
  	or $\alpha_{31}$ (right) shown by red dots.
  }
  \label{fig:Fig_A21_DEL}
\end{figure}

\begin{figure}[t]
  \centerline{
  \includegraphics[width=7cm]{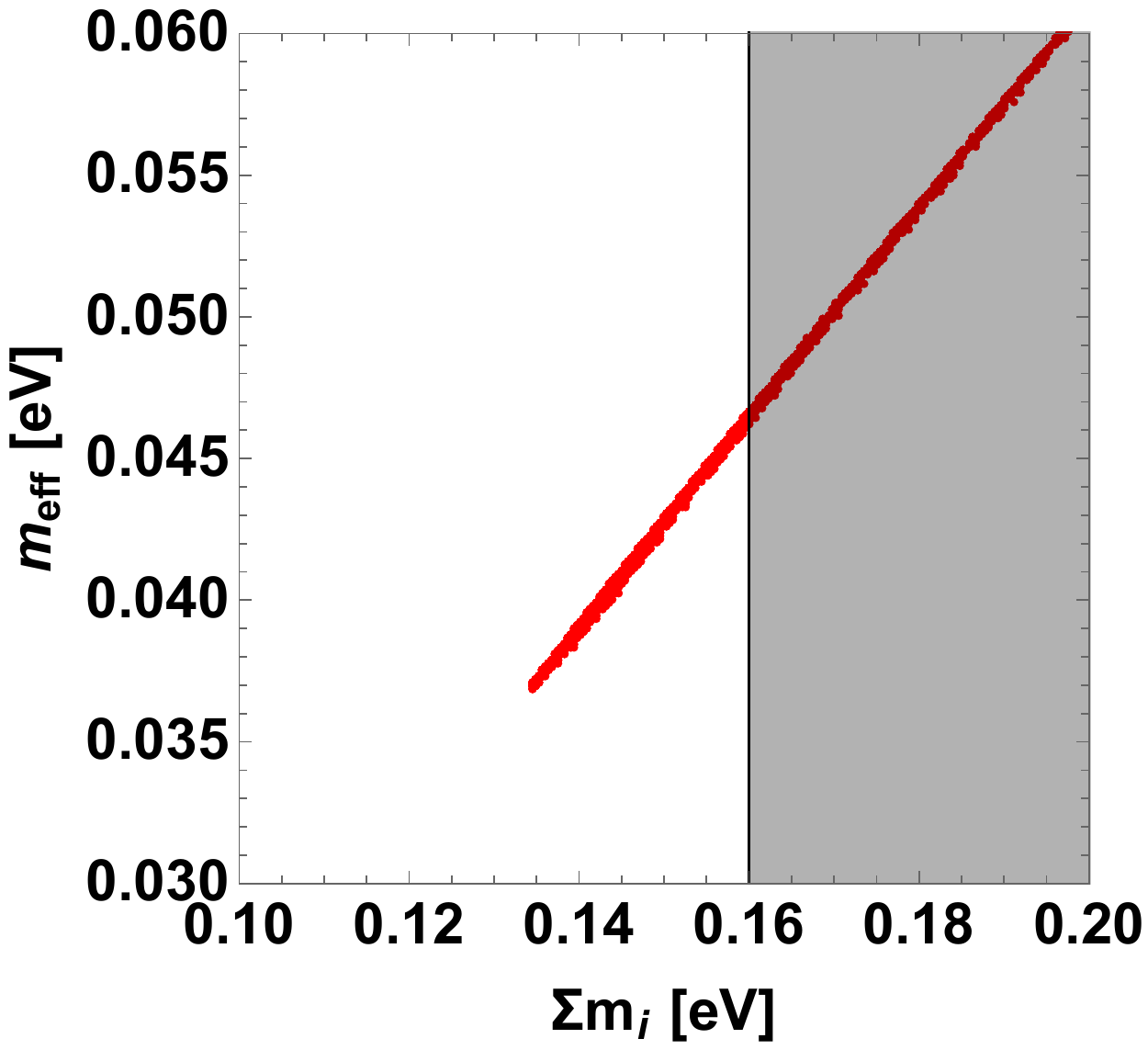}%
  }%
  \vspace{-2ex}
  \caption{
  	Predicted range of $m_{\rm eff}$ and $\sum m_i$ shown by red dots.
  	The colored region is excluded by the cosmological bound on $\sum m_i$ in Eq.~(\ref{eq:sum_mi}).
  }
  \label{fig:FIG_summi_meff}
\end{figure}

Second, we consider the predictions on the CP violating phases,
one Dirac phase $\delta_{\rm CP}$ and two Majorana phases $\alpha_{21}$ and $\alpha_{31}$,
in the neutrino mixing matrix.

In Fig.~\ref{fig:Fig_S23_DEL} we show the predicted range of $\delta_{\rm CP}$ and $\sin^2 \theta_{23}$.
It is seen that the Dirac phase is predicted as
$\delta_{\rm CP} \simeq \pm 0.5~\pi$.  This is one of the most important predictions of the model.
The recent result from the T2K experiment together with the reactor neutrino experiments
favors $\delta_{\rm CP} \simeq - 0.5 \, \pi$~\cite{Abe:2019vii}.

The Majorana phases are also predicted in the limited range as shown in Fig.~\ref{fig:Fig_A21_DEL}.
It is found that there are two possible regions.  One is
$\alpha_{21} \simeq 0 + \epsilon$ and $\alpha_{31} \simeq \pi + \epsilon$ 
for $\delta_{\rm CP} \simeq - 0.5 \pi$, where $\epsilon$ denotes some small value with $\epsilon \ll 1$.
The other is $\alpha_{21} \simeq 2 \pi - \epsilon$ and $\alpha_{31} \simeq \pi - \epsilon$
for $\delta_{\rm CP} \simeq + 0.5 \pi$.
Then, the favored value $\delta_{\rm CP} \simeq - 0.5 \pi$ leads uniquely to 
$\alpha_{21} \simeq 0 + \epsilon$ and $\alpha_{31} \simeq \pi + \epsilon$.

Finally, we discuss the impact on the neutrinoless double beta decay.
(See, for example, a review~\cite{Pas:2015eia}.)
The decay is characterized by the so-called effective mass of neutrinos defined by
\begin{align}
    m_{\rm eff} &
    = \left| \sum_{i=1}^3 U_{e i}^2 m_i \right| \,,
\end{align}
where $U_{e i}$ denotes the element of the neutrino mixing matrix.
The predicted range of $m_{\rm eff}$ is
\begin{align}
    0.037~\mbox{eV} \le m_{\rm eff} \le 0.047~\mbox{eV} \,,
\end{align}
where the upper bound comes from Eq.~(\ref{eq:sum_mi}).
See Fig.~\ref{fig:FIG_summi_meff}.
Notice that the most stringent bound on $m_{\rm eff}$ at present comes from
KamLAND-Zen experiment~\cite{KamLAND-Zen:2016pfg} as
$m_{\rm eff} \le (0.061$--$0.165)$~eV.  
Interestingly, the forthcoming KamLAND-Zen 800 experiment will explore the above range
which is one of the crucial tests of the model.

\section{Conclusions}
We have constructed the lepton flavor model based on the modular $A_4$ symmetry
which can be consistent with the neutrino oscillation data in the large volume limit 
of the torus compactification.  Although the limit leads to the simplified structure
of the modular form,  the observed mixing pattern can be reproduced by introducing
the VEV of the extra gauge-singlet field.
Such a VEV can also explain the origin of the $\mu$ term of the Higgs fields
as well as the Majorana masses for right-handed neutrinos.

The model can be consistent with the observed data only in the NH case.
The sum of neutrino masses is larger than 0.13~eV, which will be probed by
the forthcoming cosmological observations.
The mixing angle $\theta_{23}$ should be large as
$\sin^2 \theta_{23} \ge 0.58$ which will be tested by the accelerator neutrino experiments
in near future.
The CP violating phases are predicted in two cases:
One is that the Dirac phase is $\delta_{\rm CP} \simeq - 0.5 \pi$,
and the Majorana phases are $\alpha_{21} \simeq 0$ and $\alpha_{31} \simeq \pi$.  
The other is $\delta_{\rm CP} \simeq + 0.5 \pi$,
$\alpha_{21} \simeq 2 \pi$, and $\alpha_{31} \simeq \pi$.  
The present T2K experiment favors the former case.
The effective mass in the neutrinoless double beta decay is predicted 
in the range $m_{\rm eff} = 0.037$--$0.047$~eV, which will be explored
by the experiments like KamLAND-Zen 800.

\section*{Acknowledgments}
The work of T.A. was partially supported by JSPS KAKENHI Grant Numbers 17K05410, 18H03708, 19H05097, and 20H01898. The work of T.Y. was supported by the Sasakawa Scientific Research Grant from The Japan Science Society No. 2020-2024.


\clearpage
\begin{thebibliography}{100}
\bibitem{Altarelli:2010gt}
G.~Altarelli and F.~Feruglio,
Rev. Mod. Phys. \textbf{82} (2010), 2701-2729
[arXiv:1002.0211 [hep-ph]].

\bibitem{Ishimori:2010au}
H.~Ishimori, T.~Kobayashi, H.~Ohki, Y.~Shimizu, H.~Okada and M.~Tanimoto,
Prog. Theor. Phys. Suppl. \textbf{183} (2010), 1-163
[arXiv:1003.3552 [hep-th]].

\bibitem{Hernandez:2012ra}
D.~Hernandez and A.~Y.~Smirnov,
Phys. Rev. D \textbf{86} (2012), 053014
[arXiv:1204.0445 [hep-ph]].

\bibitem{King:2013eh}
S.~F.~King and C.~Luhn,
Rept. Prog. Phys. \textbf{76} (2013), 056201
[arXiv:1301.1340 [hep-ph]].

\bibitem{Petcov:2017ggy}
S.~T.~Petcov,
Eur. Phys. J. C \textbf{78} (2018) no.9, 709
[arXiv:1711.10806 [hep-ph]].

\bibitem{Feruglio:2017spp}
F.~Feruglio,
[arXiv:1706.08749 [hep-ph]].

\bibitem{Kobayashi:2018vbk}
T.~Kobayashi, K.~Tanaka and T.~H.~Tatsuishi,
Phys. Rev. D \textbf{98} (2018) no.1, 016004
[arXiv:1803.10391 [hep-ph]].

\bibitem{Penedo:2018nmg}
J.~T.~Penedo and S.~T.~Petcov,
Nucl. Phys. B \textbf{939} (2019), 292-307
[arXiv:1806.11040 [hep-ph]].

\bibitem{Criado:2018thu}
J.~C.~Criado and F.~Feruglio,
SciPost Phys. \textbf{5} (2018) no.5, 042
[arXiv:1807.01125 [hep-ph]].

\bibitem{Kobayashi:2018scp}
T.~Kobayashi, N.~Omoto, Y.~Shimizu, K.~Takagi, M.~Tanimoto and T.~H.~Tatsuishi,
JHEP \textbf{11} (2018), 196
[arXiv:1808.03012 [hep-ph]].

\bibitem{Novichkov:2018ovf}
P.~P.~Novichkov, J.~T.~Penedo, S.~T.~Petcov and A.~V.~Titov,
JHEP \textbf{04} (2019), 005
[arXiv:1811.04933 [hep-ph]].

\bibitem{Novichkov:2018nkm}
P.~P.~Novichkov, J.~T.~Penedo, S.~T.~Petcov and A.~V.~Titov,
JHEP \textbf{04} (2019), 174
[arXiv:1812.02158 [hep-ph]].

\bibitem{deAnda:2018ecu}
F.~J.~de Anda, S.~F.~King and E.~Perdomo,
Phys. Rev. D \textbf{101} (2020) no.1, 015028
[arXiv:1812.05620 [hep-ph]].

\bibitem{Okada:2018yrn}
H.~Okada and M.~Tanimoto,
Phys. Lett. B \textbf{791} (2019), 54-61
[arXiv:1812.09677 [hep-ph]].

\bibitem{Kobayashi:2018wkl}
T.~Kobayashi, Y.~Shimizu, K.~Takagi, M.~Tanimoto, T.~H.~Tatsuishi and H.~Uchida,
Phys. Lett. B \textbf{794} (2019), 114-121
[arXiv:1812.11072 [hep-ph]].

\bibitem{Novichkov:2018yse}
P.~P.~Novichkov, S.~T.~Petcov and M.~Tanimoto,
Phys. Lett. B \textbf{793} (2019), 247-258
[arXiv:1812.11289 [hep-ph]].

\bibitem{Ding:2019xna}
G.~J.~Ding, S.~F.~King and X.~G.~Liu,
Phys. Rev. D \textbf{100} (2019) no.11, 115005
[arXiv:1903.12588 [hep-ph]].

\bibitem{Nomura:2019jxj}
T.~Nomura and H.~Okada,
Phys. Lett. B \textbf{797} (2019), 134799
[arXiv:1904.03937 [hep-ph]].

\bibitem{Novichkov:2019sqv}
P.~P.~Novichkov, J.~T.~Penedo, S.~T.~Petcov and A.~V.~Titov,
JHEP \textbf{07} (2019), 165
[arXiv:1905.11970 [hep-ph]].

\bibitem{deMedeirosVarzielas:2019cyj}
I.~de Medeiros Varzielas, S.~F.~King and Y.~L.~Zhou,
Phys. Rev. D \textbf{101} (2020) no.5, 055033
[arXiv:1906.02208 [hep-ph]].

\bibitem{Liu:2019khw}
X.~G.~Liu and G.~J.~Ding,
JHEP \textbf{08} (2019), 134
[arXiv:1907.01488 [hep-ph]].

\bibitem{Okada:2019xqk}
H.~Okada and Y.~Orikasa,
Phys. Rev. D \textbf{100} (2019) no.11, 115037
[arXiv:1907.04716 [hep-ph]].

\bibitem{Kobayashi:2019mna}
T.~Kobayashi, Y.~Shimizu, K.~Takagi, M.~Tanimoto and T.~H.~Tatsuishi,
JHEP \textbf{02} (2020), 097
[arXiv:1907.09141 [hep-ph]].

\bibitem{Ding:2019zxk}
G.~J.~Ding, S.~F.~King and X.~G.~Liu,
JHEP \textbf{09} (2019), 074
[arXiv:1907.11714 [hep-ph]].

\bibitem{King:2019vhv}
S.~F.~King and Y.~L.~Zhou,
Phys. Rev. D \textbf{101} (2020) no.1, 015001
[arXiv:1908.02770 [hep-ph]].

\bibitem{Nomura:2019lnr}
T.~Nomura, H.~Okada and O.~Popov,
Phys. Lett. B \textbf{803} (2020), 135294
[arXiv:1908.07457 [hep-ph]].

\bibitem{Criado:2019tzk}
J.~C.~Criado, F.~Feruglio and S.~J.~D.~King,
JHEP \textbf{02} (2020), 001
[arXiv:1908.11867 [hep-ph]].

\bibitem{Kobayashi:2019xvz}
T.~Kobayashi, Y.~Shimizu, K.~Takagi, M.~Tanimoto and T.~H.~Tatsuishi,
Phys. Rev. D \textbf{100} (2019) no.11, 115045
[erratum: Phys. Rev. D \textbf{101} (2020) no.3, 039904]
[arXiv:1909.05139 [hep-ph]].

\bibitem{Asaka:2019vev}
T.~Asaka, Y.~Heo, T.~H.~Tatsuishi and T.~Yoshida,
JHEP \textbf{01} (2020), 144
[arXiv:1909.06520 [hep-ph]].

\bibitem{Chen:2019ewa}
M.~C.~Chen, S.~Ramos-Sánchez and M.~Ratz,
Phys. Lett. B \textbf{801} (2020), 135153
[arXiv:1909.06910 [hep-ph]].

\bibitem{Gui-JunDing:2019wap}
G.~J.~Ding, S.~F.~King, X.~G.~Liu and J.~N.~Lu,
JHEP \textbf{12} (2019), 030
[arXiv:1910.03460 [hep-ph]].

\bibitem{Zhang:2019ngf}
D.~Zhang,
Nucl. Phys. B \textbf{952} (2020), 114935
[arXiv:1910.07869 [hep-ph]].

\bibitem{Wang:2019ovr}
X.~Wang and S.~Zhou,
JHEP \textbf{05} (2020), 017
[arXiv:1910.09473 [hep-ph]].

\bibitem{Kobayashi:2019uyt}
T.~Kobayashi, Y.~Shimizu, K.~Takagi, M.~Tanimoto, T.~H.~Tatsuishi and H.~Uchida,
Phys. Rev. D \textbf{101} (2020) no.5, 055046
[arXiv:1910.11553 [hep-ph]].

\bibitem{Nomura:2019xsb}
T.~Nomura, H.~Okada and S.~Patra,
[arXiv:1912.00379 [hep-ph]].

\bibitem{Kobayashi:2019gtp}
T.~Kobayashi, T.~Nomura and T.~Shimomura,
Phys. Rev. D \textbf{102} (2020) no.3, 035019
[arXiv:1912.00637 [hep-ph]].

\bibitem{Lu:2019vgm}
J.~N.~Lu, X.~G.~Liu and G.~J.~Ding,
Phys. Rev. D \textbf{101} (2020) no.11, 115020
[arXiv:1912.07573 [hep-ph]].

\bibitem{Wang:2019xbo}
X.~Wang,
Nucl. Phys. B \textbf{957} (2020), 115105
[arXiv:1912.13284 [hep-ph]].

\bibitem{Okada:2020dmb}
H.~Okada and Y.~Shoji,
[arXiv:2003.13219 [hep-ph]].

\bibitem{Ding:2020yen}
G.~J.~Ding and F.~Feruglio,
JHEP \textbf{06} (2020), 134
[arXiv:2003.13448 [hep-ph]].


\bibitem{Kaplunovsky:1995jw}
V.~Kaplunovsky and J.~Louis,
Nucl. Phys. B \textbf{444} (1995), 191-244
[arXiv:hep-th/9502077 [hep-th]].

\bibitem{Antoniadis:1994hg}
I.~Antoniadis, E.~Gava, K.~S.~Narain and T.~R.~Taylor,
Nucl. Phys. B \textbf{432} (1994), 187-204
[arXiv:hep-th/9405024 [hep-th]].

\bibitem{Minkowski:1977sc} 
  P.~Minkowski,
  Phys.\ Lett.\  {\bf 67B}, 421 (1977);
%
\bibitem{Yanagida:1979ws} 
  T. Yanagida, in Proceedings of the Workshop on Unified Theory and Baryon Number of the Universe, edited by O. Sawada and A. Sugamoto (KEK, Tsukuba, Ibaraki 305- 0801 Japan, 1979), p. 95;
%
 %
  T.~Yanagida,
  Prog.\ Theor.\ Phys.\  {\bf 64}, 1103 (1980).
%
\bibitem{Gell-Mann:1979ws}  
M. Gell-Mann, P. Ramond, and R. Slansky, in Supergravity,
edited by P. van Niewwenhuizen and D. Freedman
(North Holland, Amsterdam, 1979);
%
  P.~Ramond,
  hep-ph/9809459;
 %
\bibitem{Glashow:1980wc}  
S. L. Glashow, in Proc. of the Cargése Summer Institute on Quarks and Leptons, Cargése, July 9-29, 1979, edited by M. Lévy et al. (Plenum, New York, 1980), p. 707;
%
\bibitem{Mohapatra:1979ia}
  R.~N.~Mohapatra and G.~Senjanovic,
  Phys.\ Rev.\ Lett.\  {\bf 44} (1980) 912.
  
\bibitem{AHY2}
  T.~Asaka, Y.~Heo and T.~Yoshida, in preparation.
  
\bibitem{nufit}
  NuFIT~5.0 (2020) [http://www.nu-fit.org/?q=node/166],
  I.~Esteban, M.~C.~Gonzalez-Garcia, M.~Maltoni, T.~Schwetz and A.~Zhou,
[arXiv:2007.14792 [hep-ph]].

\bibitem{Aghanim:2018eyx}
N.~Aghanim \textit{et al.} [Planck],
Astron. Astrophys. \textbf{641} (2020), A6
[arXiv:1807.06209 [astro-ph.CO]].
  
\bibitem{Abe:2019vii}
K.~Abe \textit{et al.} [T2K],
Nature \textbf{580} (2020) no.7803, 339-344
[arXiv:1910.03887 [hep-ex]].


\bibitem{Pas:2015eia}
H.~Päs and W.~Rodejohann,
New J. Phys. \textbf{17} (2015) no.11, 115010
[arXiv:1507.00170 [hep-ph]].

\bibitem{KamLAND-Zen:2016pfg}
A.~Gando \textit{et al.} [KamLAND-Zen],
Phys. Rev. Lett. \textbf{117} (2016) no.8, 082503
[arXiv:1605.02889 [hep-ex]].



\end{thebibliography}
\end{document}